%% file: main.tex
\documentclass{egpubl}
\usepackage{eurovis2022}
\usepackage{xspace}

\SpecialIssuePaper         % uncomment for final version of Computer Graphics Forum, special issue
\usepackage[T1]{fontenc}
\usepackage{dfadobe}  

\usepackage{cite}  % comment out for biblatex with backend=biber
\BibtexOrBiblatex
\electronicVersion
\PrintedOrElectronic
\ifpdf \usepackage[pdftex]{graphicx} \pdfcompresslevel=9
\else \usepackage[dvips]{graphicx} \fi

\usepackage{egweblnk}
\newcommand{\name}{Urban~Rhapsody\xspace}
\newcommand{\myparagraph}[1]{\noindent\textbf{#1}\xspace}

 \title[Urban Rhapsody: Large-scale exploration of urban soundscapes]%
      {Urban Rhapsody: Large-scale exploration of urban soundscapes}

\author[Rulff et al.]
{\parbox{\textwidth}{
    \centering 
    Joao Rulff$^1$, 
    Fabio Miranda$^2$, 
    Maryam Hosseini$^1$,
    Marcos Lage$^3$,
    Mark Cartwright$^4$,
    Graham Dove$^1$,
    Juan Bello$^1$,
    Claudio T. Silva$^1$   
    }\\
{\parbox{\textwidth}{\centering 
        $^1$New York University,
        $^2$University of Illinois at Chicago, 
        $^3$Universidade Federal Fluminense, 
        $^4$New Jersey Institute of Technology
       } 
}
}
\begin{document}

\teaser{
 \includegraphics[width=\linewidth]{figures/teaser2-final_compressed.pdf}
 \centering
 \caption{We use \name to assess after-hour construction in New York City, first selecting audio recordings captured by sensors deployed around Broadway. \name allows users to query using an audio sample, and drill down to days containing similar audios~(a). Using the interactions provided by the tool, we are able to create classification models according to a user's perception of the soundscape (b,c), and then use these models to classify the entire data set and look for unusual events (d,e).}
\label{fig:teaser}
}

\maketitle
%-------------------------------------------------------------------------
\begin{abstract}
Noise is one of the primary quality-of-life issues in urban environments. In addition to annoyance, noise negatively impacts public health and educational performance.
While low-cost sensors can be deployed to monitor ambient noise levels at high temporal resolutions, the amount of data they produce and the complexity of these data pose significant analytical challenges. 
One way to address these challenges is through machine listening techniques, which are used to extract features in attempts to classify the source of noise and understand temporal patterns of a city's noise situation. 
However, the overwhelming number of noise sources in the urban environment and the scarcity of labeled data makes it nearly impossible to create classification models with large enough vocabularies that capture the true dynamism of urban soundscapes.
In this paper, we first identify a set of requirements in the yet unexplored domain of urban soundscape exploration.
To satisfy the requirements and tackle the identified challenges, we propose Urban Rhapsody, a framework that combines state-of-the-art audio representation, machine learning and visual analytics to allow users to interactively create classification models, understand noise patterns of a city, and quickly retrieve and label audio excerpts in order to create a large high-precision annotated database of urban sound recordings. 
We demonstrate the tool’s utility through case studies performed by domain experts using data generated over the five-year deployment of a one-of-a-kind sensor network in New York City.
%-------------------------------------------------------------------------
%  ACM CCS 1998
%  (see https://www.acm.org/publications/computing-classification-system/1998)
% \begin{classification} % according to https://www.acm.org/publications/computing-classification-system/1998
% \CCScat{Computer Graphics}{I.3.3}{Picture/Image Generation}{Line and curve generation}
% \end{classification}
%-------------------------------------------------------------------------
%  ACM CCS 2012 (see https://www.acm.org/publications/class-2012)
   
%The tool at \url{http://dl.acm.org/ccs.cfm} can be used to generate
% CCS codes.
%Example:


\end{abstract}

%% sections
\input{sections/1-introduction}
\input{sections/2-background}
\input{sections/3-related}
\input{sections/4-sonyc}
\input{sections/5-requirements}
\input{sections/6-system}
\input{sections/7-cases}
\input{sections/8-conclusion}
\input{sections/acknoledgements}

% bibtex
\bibliographystyle{eg-alpha-doi}  
\bibliography{references}        

% biblatex with biber
% \printbibliography                

%-------------------------------------------------------------------------

\end{document}

%% file: sections/1-introduction.tex
%-------------------------------------------------------------------------
\section{Introduction}
\maketitle
City soundscapes represent a rich source of information about urban systems, such as transportation, civil construction, and social activity. Low-cost sensors can be used to capture aspects of this acoustic environment, and computational methods for large-scale data analysis offer new approaches to characterizing the different contributing sources. Such understanding offers insight into how a city behaves through space and time (e.g., \emph{"what are the typical sounds in a neighborhood during the night?"}), and can help in tackling various urban problems such as noise pollution.
The research reported here was undertaken in partnership with researchers from one such sensing initiative, the Sounds of New York City (SONYC) project~\cite{bello_sonyc_2019}, who have developed and deployed low-cost sensors to measure and stream real-time sound pressure level (SPL) and audio data. 
To date, more than fifty sensors have been deployed throughout New York City (NYC), collecting data for over five years (in total, more than 60~TB). 
To meaningfully understand this data, the project's researchers are developing new machine listening models that 1)~extract audio embeddings and 2)~classify these sounds based on a set of predefined labels. 
However, these tasks pose several challenges that impede even state-of-the-art models' effectiveness in capturing the urban soundscape's dynamism.
First, audio is complex, a recording typically captures different sound sources (e.g., dogs barking and people talking) simultaneously. Second, sound events are transient (e.g., a honking car horn) but in aggregation can last for hours (e.g., car engines on a busy highway). Third, audio has a temporal aspect, and so unlike  images or words, sounds do not have a straightforward pictorial representation, limiting our ability to quickly review a large collection of recordings in parallel. Hundreds of images can be reviewed at the same time, with objects identified in minutes. However, looking for patterns or events in a large collection of audio data often requires listening to hours of individual recordings one after another. Analyzing audio data is time-consuming and hard to scale. This calls for novel techniques and visualization interfaces to facilitate the process, leveraging human expertise.

Motivated by these challenges and the need to gain new insights into the soundscape of the city, we introduce \name, a framework for the interactive visual analysis of large collections of urban acoustic data.
Using recent advances in machine listening to generate audio representations, \name allows analysts to create a visual representation of the soundscape across different ranges of temporal and geographical granularity. 
We adopt a human-in-the-loop approach that enables users to interactively label data points, create new classification models based on their expertise of the soundscape, and assess the performance of audio classification tasks.
Finally, because noise patterns might happen at different scales (minutes, days, months, etc.) in the urban environment, we employ a multilevel visualization scheme. 
Using case studies that demonstrate the utility of \name, we showcase support for fast exploration of similar sounds or concepts, assessment of classification model outputs in different scenarios, geographical and temporal understanding of the embedding space, and summarization of soundscapes by key representative audio frames. 
Previous approaches to these challenges were either applied in a different context ~\cite{dema2017collaborative}, or constrained to the analysis of sound pressure level (SPL) data~\cite{miranda_time_2018}, painting an incomplete picture regarding urban noise problems ~\cite{zheng2014diagnosing}. \name is the first visual analytics framework that enables a comprehensive analysis of urban acoustic environments, going beyond time series to leverage a unique audio data set that enables a more comprehensive analysis.
Our contributions can be summarized as follows: (1) A set of requirements, elicited in collaboration with SONYC's audio researchers, for visual exploration of large urban audio sets.   
(2) A set of visual interactions that enables users to iteratively construct audio machine learning models; (3) An interactive visual analysis framework, \name, that supports concept-based exploration of large collections of audio recordings (such as the ones generated over the five-year deployment of the SONYC sensor network). We illustrate this with two case studies set in NYC, highlighting how our approach can be useful in tackling issues that have generated intense public debate. Our framework is also available on GitHub (\url{https://github.com/VIDA-NYU/Urban-Rhapsody}).

%% file: sections/2-background.tex
\section{Background}
According to the World Health Organization, in Western Europe alone, more than 1 million healthy life-years are lost annually to environmental noise pollution~\cite{organization_burden_2011}, and in NYC, an estimated 9 out of 10 adults are exposed to excessive noise levels~\cite{neitzel_exposures_2012}.
This impacts public health~\cite{hammer_environmental_2014}, social well-being~\cite{guite_impact_2006} and quality of life ~\cite{dratva_impact_2010}, as noise increases stress, sleep disruption, annoyance and distraction~\cite{bronzaft_neighborhood_2007,organization_burden_2011,haralabidis_acute_2008,muzet_need_2002}.
To mitigate this, governments devise noise codes that typically consider SPL measurements in relation to time of the day/week and location and impose regulations that aim at mitigating the noise at the source (e.g., by erecting sound barriers around major roads or modifying building designs)~\cite{taber_technology_2007,bronzaft_noise_2010,hammer_environmental_2014}.
However, enforcing these codes is time-consuming and costly, requiring trained inspectors to be present at sites to make assessments and capture sound carefully using calibrated equipment~\cite{bello_sonyc_2019}.

Beyond this, noise pollution can be highly subjective~\cite{de_paiva_vianna_noise_2015}, and so quantitative SPL metrics may be insufficient~\cite{raimbault_ambient_2003,guastavino_etude_2003}. Because of this, there is a shift towards understanding the \emph{source} of the noise, and to consider context in people's perception of sounds~\cite{raimbault_urban_2005,van_kempen_characterizing_2014}. Such a ``soundscape approach'' ~\cite{payne_research_2009,brown_soundscapes_2010,davies_perception_2013} views the acoustic environment as composed of both positive and negative sources~\cite{brown_review_2012}.
Data gathered using SONYC sensors offers a unique opportunity to measure noise pollution quantitatively and additionally gain insights into the acoustic environment's qualitative characteristics. 
We can therefore conduct structured assessments at scale, accounting for both SPL and sound source. This raises important challenges (outlined in Section~\ref{sec:challenges}) that we seek to address in this research. \name is the first step towards allowing domain experts to better understand the soundscape of complex cities such as NYC.

%% file: sections/3-related.tex
%-------------------------------------------------------------------------
\section{Related Work}

\subsection{Urban visual analytics}
Urban areas are a major source of data that have tremendous potentials to improve policy making, enhance the lives of citizens, and pursue sustainable development.
Visualization systems have for long been an important tool for the analysis of urban data~\cite{zheng_visual_2016,8474495}.
Several approaches use urban data to study different properties of a city, such as air pollution~\cite{zheng_u-air_2013}, public utility service problems~\cite{zhang_visual_2014}, sunlight access~\cite{8283638}, land use~\cite{quercia_mining_2014}, human movement patterns~\cite{noulas_tale_2012,lenormand_human_2015,7539380}, transportation~\cite{andrienko_spatio-temporal_2008,wang_visual_2013,ferreira_visual_2013,itoh_visual_2014,zeng_visualizing_2014}, and also the relationship between these data sets~\cite{malik2012correlative, chirigati_data_2016, deng2021compass}. 
More general tools, such as ArcGIS~\cite{johnston_using_2001}, Urbane~\cite{ferreira_urbane_2015,Doraiswamy:2018:IVE:3183713.3193559}, and Vis-A-Ware~\cite{ortner_vis--ware_2016} have facilitated the use of multiple urban data sets to help inform urban planning and decision making process.

In our previous work, Time Lattice~\cite{miranda_time_2018}, we have tackled the problem of noise pollution by proposing a data structure and visual interface that allowed experts to explore a large data set composed of SPL dB measurements from SONYC sensors. We only used SPL measurements without considering that the soundscape of a city is composed of different sources and can be perceived differently by different people.
With \name, our goal is to account for the user's knowledge and perception in the exploratory process of large collections of urban sound recordings in a vocabulary-free approach, meaning that users are free to explore the soundscape according to any concept they create. To the best of our knowledge, \name is the first visual analytics system specifically designed to allow domain experts to explore a large collection of sound recordings of an urban environment.

\subsection{Environmental sound representation}

In recent years, several large audio data sets have been released that have moved the field of environmental machine listening forward \cite{gemmeke_audio_2017,fonseca_fsd50k_2020}. 
However, many audio classification tasks do not map onto the class vocabulary of these data sets and thus require additional labeling, which is time-consuming and costly. 
To address this problem, machine listening practitioners have turned to transfer learning \cite{yosinski_how_2014} in recent years, which has been shown to be effective for many audio classification tasks \cite{aytar_soundnet_2016,arandjelovic_look_2017,jansen_unsupervised_2018,kumar_knowledge_2018,cartwright_tricycle_2019,cramer_look_2019,tagliasacchi_pre-training_2020,grollmisch_analyzing_2021}. 
In transfer learning, models are typically pre-trained on large data sets using supervised \cite{aytar_soundnet_2016,hershey_cnn_2017} or self-supervised learning \cite{arandjelovic_look_2017,jansen_unsupervised_2018,kumar_knowledge_2018,cartwright_tricycle_2019,tagliasacchi_pre-training_2020}, and the knowledge acquired during pre-training is re-used for tasks where data is limited. 
A common method of re-using this knowledge is to treat the pre-trained models as feature extractors, utilizing learned latent representations (i.e., embeddings) from within the pre-trained models as the inputs to models with little or no labeled data.
Look, Listen, and Learn \cite{arandjelovic_look_2017} is one such pre-trained model whose embeddings were shown to be discriminative in several environmental audio classification tasks \cite{arandjelovic_look_2017,cramer_look_2019,cartwright_tricycle_2019,wilkinghoff_open-set_2021}. 
This model is pre-trained using self-supervision on an auxiliary task of audio-visual correspondence. 
%
% In this task, the model aims to determine whether a video image frame and a 1 second audio segment come from the same video and overlap in time. 
% Thus, it is trained without any human-labeled data and instead leverages the relationships between the auditory and visual modalities. 
In this work, we use OpenL3 \cite{cramer_look_2019}, an open-source code of Look, Listen, and Learn,  as an audio feature extractor to transform each audio recording into a series of embedding representations.

\subsection{Machine-learning-aided multimedia exploration}

Machine learning has opened a new horizon in data exploration across various fields, with numerous systems making use of the powerful capabilities it provides. For instance, Urban Mosaic~\cite{10.1145/3313831.3376399} uses deep learning representations to search for patterns in a large collection of street-level images. II-20~\cite{zahalka2020ii} allows users to generate image classifiers using novel interactions. Previous works tried to explore the semantic meaning of the features extracted by deep learning models, as they do not always map into human-understandable semantic features: Embedding Projector~\cite{smilkov_embedding_2016} and Latent Space Cartography~\cite{liu_latent_2019} enable the analysis of embedding spaces for multimedia data through multidimensional projections~\cite{joia_local_2011,mcinnes_umap_2018} to enable users to understand features that might be encoded in the latent representations. 

To create classification models that can recognize human-understandable features in multimedia data sets, previous approaches employ active learning frameworks leveraging the users as oracle annotators to annotate new samples the system identified as the most informative. For example, previous studies investigated the usefulness of active learning for labeling tasks~\cite{bernard2017comparing} and its application in other fields such as anomaly detection~\cite{liao2010anomaly,lin2017rclens}, commuting flow estimation~\cite{yu2015iviztrans}, and image categorization~\cite{zahalka2020ii}.
These approaches often guide the user on choosing the next subsets of the data to label next to improve the performance of the model, which, in our case, can limit the user in applying their previous understanding of the soundscape to label the concepts~\cite{nadj2020power}.
Our proposal leverages a set of techniques presented in previous works to enable users to better understand the spatiotemporal distribution of events in acoustic recordings while accounting for their knowledge of the soundscape to build concept-based classification models to gain insights into the dynamics of the urban environment.

%% file: sections/4-sonyc.tex
\begin{figure}[t!]
    \centering
    \includegraphics[width=\linewidth]{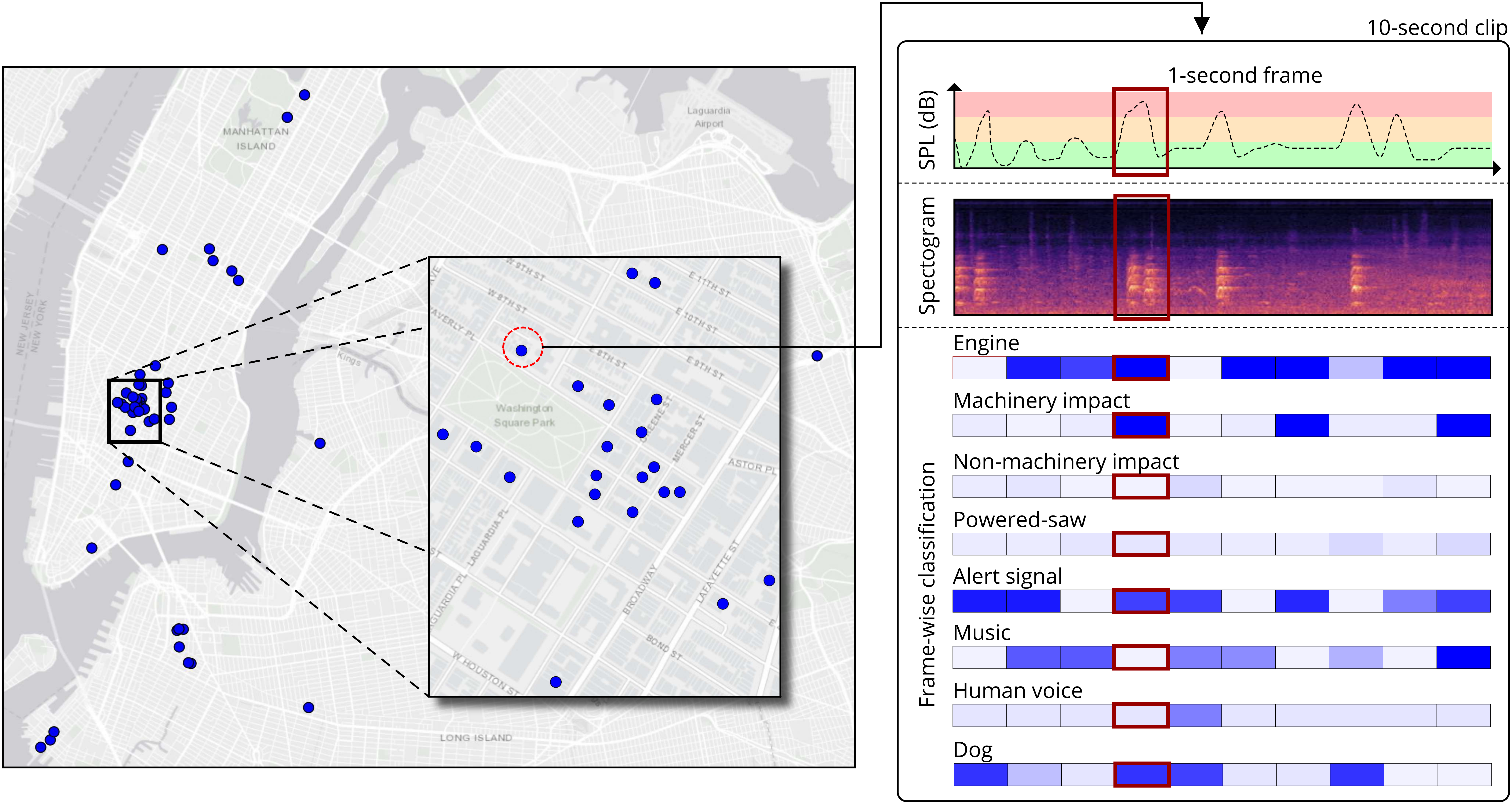}
    \caption{Spatial distribution of SONYC sensors (left) showing the coverage of the city. Right image illustrates the data from a sensor located near a park in Manhattan. Sensors record both the sound pressure level at each second (SPL dB), as well as the environmental sounds (stored as 10-second clips). For each 1-second frame in the clip (highlighted in red), we compute the classification considering user-crated prototypes. The figure shows classes following standard urban audio taxonomies.} 
    \label{fig:sonyc}
\end{figure}
\section{Sounds of New York City}
The research reported in this paper was undertaken in conjunction with audio and machine listening experts from the SONYC project~\cite{bello_sonyc_2019}, and utilizes data generated by the project's sensors. Our collaborators have background in urban science and machine listening~\cite{Doraiswamy:2018:IVE:3183713.3193559,cartwright_tricycle_2019, 10.1145/3313831.3376399, wang2021calls}. In addition, the project communicates their findings to the media~\cite{sonycnyt2020} and works closely with the NYC Dept. of Environmental Protection to understand their needs and investigate new ways of monitoring and mitigating noise pollution.

\subsection{A data set of urban sounds}
The SONYC acoustic sensor network consists of more than 50 sensors deployed around three boroughs of NYC: Manhattan, Brooklyn, and Queens. Figure~\ref{fig:sonyc} shows the spatial distribution of the sensors. These sensors are positioned 15 to 25 feet above the ground. To maintain the privacy of bystanders and prevent the recording of intelligible conversations, the sensors do not record continuously, but rather record 3 10-second recordings at random intervals within each minute of a day (i.e., for a single day, each sensor will record 720 minutes worth of 10-second audio recordings uniformly distributed throughout the 1440 minutes of the day).
As of 2021, SONYC has collected approximately 1,700,000 hours of SPL data (stored as second or millisecond resolution timeseries), and 877,000 hours of recorded audio.
To extract a discriminative, lower-dimensional representation of each 10-second recording, we employ OpenL3 trained on an environmental sound subset of AudioSet~\cite{cramer_look_2019}. OpenL3 is an open-source library for computing deep audio embeddings, developed by researchers from SONYC, and its design choices were informed by the need to classify sounds from urban environments.
For each 10-second recording, we use a hop size and window size of 1.0 second (with centered windows) to compute 10 512-dimensional feature vectors. This produces a feature vector that coarsely captures the \emph{general} acoustic aspects of the environment.

\subsection{Challenges}
\label{sec:challenges}
The complexity of the urban environment brings several challenges when it comes to analyzing and extracting insights from urban sound data, especially considering such a large data set as the one captured by SONYC.

\myparagraph{Sound representation.}
In complex environments such as cities, many sound classes seem quite similar, such as car alarms and sirens, but are distinct in the noise code and should be treated as such. On top of that, when handling sounds from cities, the acoustic environment changes by location and by time within seasonal cycles. 
As a self-supervised method, OpenL3 does not need human-generated labels to be trained, while still providing good sensitivity to different urban sounds. However, it falls short of properly accounting for \emph{all} of the complexity of the soundscape of a city.

\myparagraph{Mixture of sounds.}
Unlike images, where visual objects are opaque, sound objects are conceptually \emph{transparent}, meaning that multiple objects (sound sources) can have energy at the same frequency~\cite{wyse_audio_2017}. This is especially true in an environment as complex as cities, where sounds are emitted from multiple sources, creating a soundscape that, albeit quite characteristic, is very difficult to parse and understand. In other words, in a city, at any given instant in time, a sound recording might have a mixture of background (e.g., bird songs, dog barks) and foreground sounds (e.g., engine, party, sirens).

\myparagraph{Sound exploration.}
Again unlike images, there is no clear pictorial representation of audio data. This gap between audio data and visual representation is challenging when building visual analytics systems. Visual objects are \emph{opaque} (a given pixel in a visual image corresponds to only one object), whereas sound objects are \emph{transparent} (multiple objects can have energy at the same frequency).
Sounds are therefore serial objects: when assessing an image, we can visually \emph{scan} it to identify each visual object in the scene, creating a visual map of the objects that can help us fully understand the scene. Sounds only exist at one moment in time; once the moment is gone, the sound is also gone. In other words, a user can only observe a sound one moment at a time, unlike images where we can observe multiple objects at a time. 
In spectrograms representation, similar neighboring pixels cannot be assumed to belong to the same object (i.e., frequencies are non-locally distributed on the spectrogram~\cite{wyse_audio_2017}).
As we can notice, creating visual representations of sounds is a challenge, specifically considering a scenario with multiple sound sources, such as urban soundscapes.

\myparagraph{Sound labeling and classification.}
Although previously proposed classifiers provide a reasonable link between embeddings and human-understandable vocabulary, their class vocabularies are limited, providing a narrow view of the rich and varied soundscape of the city, which is comprised of numerous types of sound events.
Furthermore, manually labeling sound data to be used as groundtruth for model training is a laborious process. As previously mentioned, sounds are serial objects where the user needs to listen to one at a time, limiting the number of audio files that can be labelled in a short period of time. Purely automated mechanisms, however, are prone to misclassifications given the complexity of soundscapes.

\myparagraph{Data size.}
Over the past five years, SONYC has generated more than 60~TB of data, including high-resolution SPL timeseries and audio recordings.
If we consider the embeddings computed with OpenL3, we have 86,400 feature vectors with size 512 (177~MB in total) \emph{per sensor per day}.
Any visualization system must properly handle such data size to be interactive~\cite{liu_effects_2014}, either by sampling, filtering or aggregating the data.

%% file: sections/5-requirements.tex
\section{System Requirements}
\label{sec:requirements}
In our collaboration with machine listening researchers, over the course of two years in the context of the SONYC project, we established a set of requirements for a visual analytics tool to facilitate their analysis workflows. We then validated the working system through interactive demo sessions exploring a number of potential use cases.
Underlying our work is the necessity to account for user knowledge when exploring the urban soundscape for different concepts.
During these meetings, we identified the following main tasks that the experts desire to perform with the tool: 1) Select and listen to sound recordings from a set of sensors, considering different days of the week and time ranges; 2) Considering a query audio, quickly identify a set of possible similar sounds throughout a long period; 3) Create and refine classification models that allow for searching of complex sound scenes; 4) Assess classification performance interactively.
To accomplish the listed tasks, we identified the following system requirements:

\myparagraph{[R1] Interactive identification and labeling of similar sounds.}
Given the highly complex acoustic environment we observe in cities, audio representations cannot encode specific audio events that users might be interested in. Moreover, the high-dimensional nature of audio representations makes it hard to visually analyze such data, making multidimensional projection techniques a standard in this process. However, in many cases, user-perceived similarities between sets of audio frames (i.e., a one-second slice of the ten-second audio snippet) might not be represented in the selected projection technique, e.g., similar frames are far apart in the projected space (low-dimensional space), making it harder for users to find similar audio frames. Hence, finding similar audio frames based on user's perception is one of the requirements of the \name framework.

\myparagraph{[R2] Projection steering based on user perception.}
When exploring audio embeddings extracted from urban recordings through multidimensional projections, we often recognize clusters that do not represent the user's perception of the soundscape. 
Based on the user's understanding of the data set expressed through labeled points, the system should provide the capability of producing new projections that better encode the user's perception.

\myparagraph{[R3] Iterative creation of classification models.}
Considering that current machine listening models present certain limitations, the system should provide the capability to iteratively create new classification models based on the data points labeled by the user (and, therefore, the user's perception of the soundscape).
The system should also support assessing the evolution of the model's convergence through successive iterations.

\myparagraph{[R4] Local and global sound perspectives.}
%
% \joao{Can someone review this paragraph?}
Audio embeddings might possess certain characteristics that only become clear when analyzed locally or globally. Then, it is important for the user to assess their local characteristics and to relate one sound to its immediate neighborhood or distant clusters.

\myparagraph{[R5] Match between audio and visual representations.}
Visualizing audio files in the frequency domain is important for the user when assessing the accuracy of both the embeddings and classifications. For instance, two sounds might have very similar spectro-temporal patterns and classifications but completely different embeddings; it is important, therefore, to further assess and create hypotheses on what led to these different outputs.

\myparagraph{[R6] Support interactive query times.}
The system should support interactive queries to enable the easy and quick labeling of data points and the creation of classification models.

%% file: sections/6-system.tex
\begin{figure*}[t!] 
    \centering
    \includegraphics[width=0.95\textwidth]{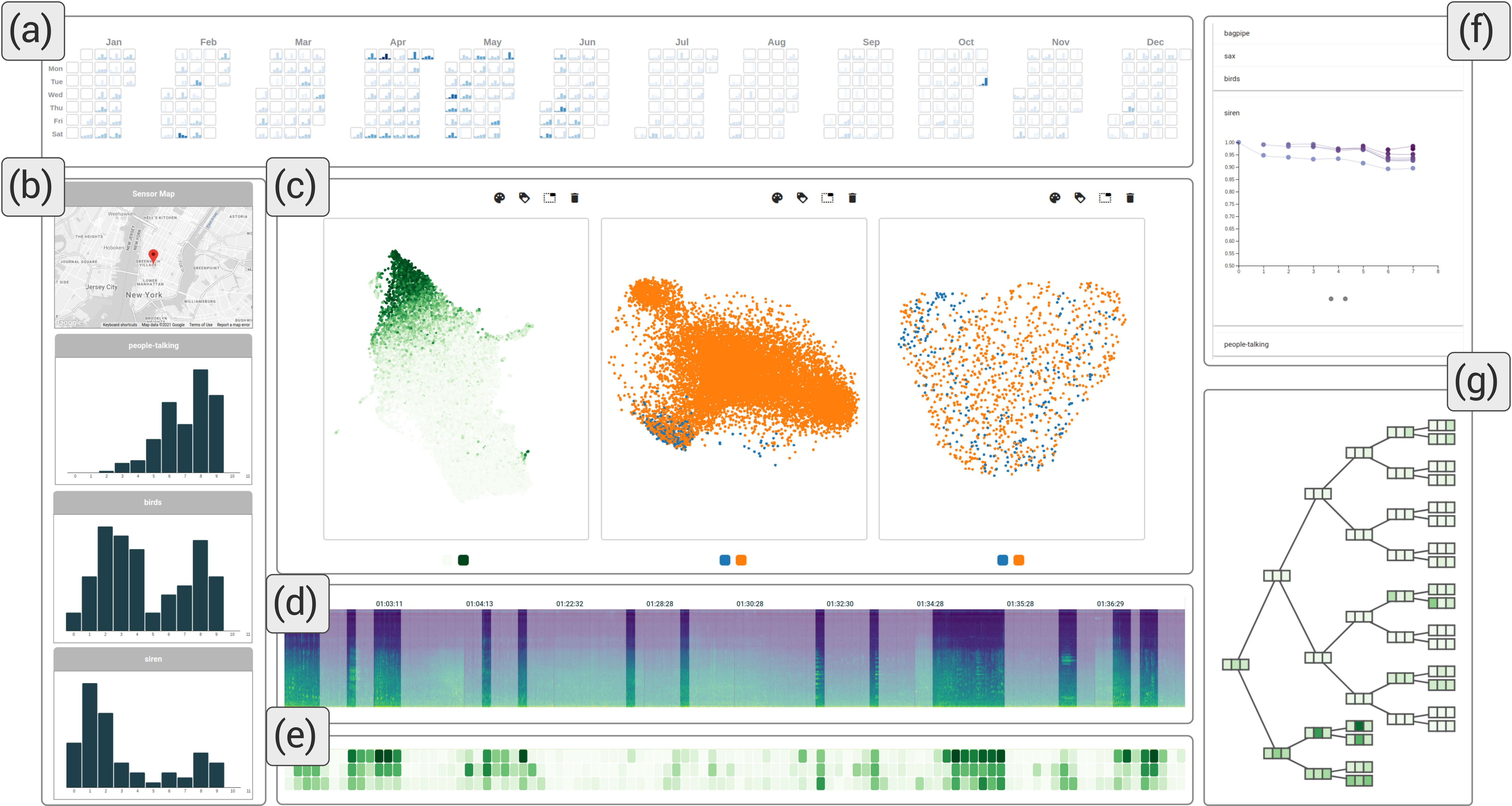}
    \caption{The \name system visual interface: (a)~Calendar View; (b)~Sensor Map and Distribution View; (c)~Day View (projections); (d)~Focused View (spectrograms); (e) Frame Classification View; (f) Model Summary; (g) Mixture Explorer.}
    \label{fig:visual_interface}
\end{figure*}

\section{\name}
To satisfy the previous requirements, we introduce \name. A visual analytics tool able to provide a human-centered exploration of the urban soundscape using prototypes created on the fly through different interaction mechanisms.
Our description of the framework is broadly divided into three parts. First, we describe our approach to generate classification models (or \emph{prototypes}) of different \emph{concepts} denoting complex urban sound scenes. Second, we describe the different components of \name's visual interface (also see accompanying video), followed by a discussion of its architecture and implementation.

%%%%%%%%%%%%%%%%%%%%%%%%%%%%%%%%%%%%%%%%%%%%%
\subsection{Prototype-based interaction}
%%%%%%%%%%%%%%%%%%%%%%%%%%%%%%%%%%%%%%%%%%%%%

In \name, we would like to support the search for audio events based on concepts and not only based on a single audio event.
Here, we use the term \emph{concept} to refer to an abstract idea or a general representation of a category in mind, such as ``crowded street'', which can be perceived differently by people. In one of our case studies, we describe a case where the user keeps refining their concept of construction while annotating new sounds that together compose the full picture of a construction. To allow for this kind of search, we define \emph{prototypes}, a structure composed of a classification model and a set of representative audio frames that defines a user's understanding of a concept.

The classification model learns how to distinguish between the audio frames that are part of a given concept according to the user's perception represented by annotations made during the interaction process with the system. Once the user starts labeling a specific concept in \name, they can generate a new classification model that will be trained using annotated frames as input. Since our goal is to find occurrences of specific concepts in our data set, we should train this model with a diverse enough sample of the data so it can generalize well to different scenarios. Given this constraint, we train our model to distinguish between two labels: positive (frame is part of the concept) and negative (not part of the concept). For positive labels, we use all the frames annotated as the concept we are interested in. For negative labels, we use frames explicitly annotated as not being part of a concept and a random sample of all frames in our data set twice as big as our set of positive-labeled frames. The classifier we train in \name is based on the classic random forest algorithm using a standard parameter setting for audio classification~\cite{wang2019active}. However, any classification model capable of outputting a likelihood score of a data point belonging to a class can be used in \name. In this version, the likelihood function is calculated as the average prediction score across the trees in the forest. This interaction supports requirement \textbf{R3}.

Following \textbf{R6}, \name must be capable of providing interactive query times during the exploration process. However, the size of the data set handled by our framework blocks us from filtering interesting audio frames by scanning the entire data set and computing the prediction probability of a given model to generate our visualizations. For this reason, after every model refinement made by the user, we also calculate a set of representative audio frames that will help us sample the data set to a smaller size before filtering interesting points using the aforementioned classification model. 
We calculate representative points of a concept by selecting all the points annotated as being part of a concept by the user and running a density-based clustering algorithm on the positive-annotated frames for a concept. For each cluster, we calculate the frame closest to its centroid and add it to the set of representative frames of that concept. The representative audio frames also help the users keep track of the concept they are creating through their interaction with the system.
We enable the user to use these representative points as query input for a concept search using an approximated nearest neighbors (ANN) query.

%%%%%%%%%%%%%%%%%%%%%%%%%%%%%%%%%%%%%%%%%%%%%
\subsection{Visual interface}
%%%%%%%%%%%%%%%%%%%%%%%%%%%%%%%%%%%%%%%%%%%%%

The visual interface was designed to provide the user with the ability to browse through the entire data set, identify and annotate concepts present in audio samples, and, finally, iteratively and interactively build prototype models that generalize these concepts over the entire data set. 
Figure~\ref{fig:visual_interface} shows the different components of the visual interface. Next, we discuss the design of each visualization based on its functionality: provide easy navigation through the audio collection, enable the annotation of audio concepts, allow for the detailed inspection of individual audio samples and facilitate the evaluation of prototype models.
%

%%%%%%%%%%%%%%%%%%%%%%%%%%%%%%%%%%%%%%%%%%%%%
\myparagraph{Audio collection navigation.} 
 %%%%%%%%%%%%%%%%%%%%%%%%%%%%%%%%%%%%%%%%%%%%%
The interface implements several strategies to enable navigating through our data set. The first is the Calendar View (Figure~\ref{fig:visual_interface}(a)). This component presents a calendar of the year with each cell representing a single day. Within each cell, we can visualize a bar chart representing the distribution of frames of a specific concept during the four time slices of a day, allowing for the fast identification of the daily distribution of sounds. The bars of each cell are also colored according to the density of a specific concept in a day (more examples in a day will lead to darker blue bars). If a Calendar View cell is clicked, all the data corresponding to that specific day is loaded and in the day view (Figure~\ref{fig:visual_interface}(c)).
Using the Day View, we can visualize the audio frames through the analysis of scatterplots generated by projecting high-dimensional feature vectors (audio embeddings) into a two-dimensional space using UMAP~\cite{mcinnes_umap_2018}. Although UMAP was the projection technique used for this version of \name, given its dimensionality reduction capabilities, it is important to notice that \name is agnostic of projection technique. The adaptation of the system to better accommodate experts' needs in terms of projection techniques is trivial. 
Here, the users can horizontally stack projections in three ways: reprojecting a subset of the data available for a day (i.e., reproject specific clusters to capture local structures of the data), removing a subset of the data, and reprojecting the remaining points (useful for removing clusters representing sensor failure, for example), and, lastly, steer the projections based on frames annotated by the user using a semi-supervised dimensionality reduction algorithm \cite{szubert2019structure} that can learn a new low-dimensional space that better encodes the user's perception of the data (i.e., bringing frames with the same labels closer while keeping the different ones distant from each other), therefore supporting \textbf{R2}.

The projections in the Day View are linked and allow for selecting points through a bounding box or periods of the day. Selections update the Distribution View as well as the components designed for the individual inspection of audio samples, the Focused and the Frame Classification Views, shown in Figure~\ref{fig:visual_interface}(d, e) are described later in this section. At last, the projected points, each representing an audio frame, can be colored according to a likelihood of belonging to a concept or user annotation.
When a day is loaded, \name automatically calculates a hierarchical clustering of the points and updates the Mixture Explorer, represented in Figure~\ref{fig:visual_interface}(g) by a tree. Each node of the tree represents a cluster found by the algorithm. Each node is subdivided into subnodes, each being one concept that the user previously created. In the example presented in Figure~\ref{fig:visual_interface}(g) each subnode is representing a concept (people talking, birds, and siren from left to right) and is colored based on the average likelihood of the correspondent cluster contain the specific concept (darker green for higher likelihoods). If a node is clicked, the corresponding cluster is selected in the scatterplots and all the components of the interface are updated accordingly. For example, the node where all subnodes are darker green is where the user is more likely to find frames that contain all created concepts.
It is important to notice that hierarchical clustering is a powerful visual strategy that enables the user to explore clusters of different sizes, both locally and globally (\textbf{R4}), and gain new insights into sound mixtures by focusing its inspection on cluster where previously created concepts are more likely to be found.

%%%%%%%%%%%%%%%%%%%%%%%%%%%%%%%%%%%%%%%%%%%%%
\myparagraph{Annotation of audio concepts.}
%%%%%%%%%%%%%%%%%%%%%%%%%%%%%%%%%%%%%%%%%%%%%
%
One of the requirements elicited with domain experts is regarding the ability to annotate specific audio frames \textbf{(R1)}. To satisfy this requirement, \name provides a mechanism to annotate specific audio frames that works as follows: users can select specific frames by using the selection mechanisms provided by the scatterplots or select a cluster using the hierarchical tree. Once a selection is made, the users can click on the labeling icon on top of the scatterplot to open a dialog that will allow for the annotation of these frames with as many labels as they want (positive labels). Also, users are able to annotate frames with negative labels, to explicitly say that a selection of frames is not part of a specific concept. This will help refine the prototype models when we find false positives during the exploration process. 

%%%%%%%%%%%%%%%%%%%%%%%%%%%%%%%%%%%%%%%%%%%%%
\myparagraph{Inspection of audio samples.} 
%%%%%%%%%%%%%%%%%%%%%%%%%%%%%%%%%%%%%%%%%%%%%
To inspect details of an audio recording selected by the user during the exploration of the projections, \name contains two widgets with visualization metaphors commonly used by audio experts: the Focused View (Figure~\ref{fig:visual_interface}(d)) and the Frame Classification View (Figure~\ref{fig:visual_interface}(e)). 
The Focused View shows a spectrogram of the audio samples selected in the projection.
A spectrogram is a visual representation of the magnitude of the short-time Fourier transform, which describes the signal's energy by frequency as it varies with time.
It can be visually encoded in a heatmatrix where each cell represents the intensity of a frequency in a given time. For example, the spectrogram of an audio file containing the sound of a siren contains wave patterns.
Previous work investigated the usefulness of spectrograms in representing audio classes for humans and its performance in comparison to others standard audio visualizations~\cite{cartwright2017seeing}.
We use this representation to allow the user to compare different sounds without having to listen to multiple audio files. 
The Frame Classification View displays the likelihood of observing a concept in the audio sample. In this way, the color of each cell of the matrix represents the probability of observing different sound classes in the associated audio frame.
Finally, \name allows the user to click on the spectrogram to listen to the recording. This interaction is important to bridge the gap between the visual representation and the actual audio (\textbf{R5}).

%%%%%%%%%%%%%%%%%%%%%%%%%%%%%%%%%%%%%%%%%%%%%
\myparagraph{Evaluation of prototypes.} 
%%%%%%%%%%%%%%%%%%%%%%%%%%%%%%%%%%%%%%%%%%%%%
%
As users keep creating and refining prototypes, they can evaluate its performance by making use of several components of our interface. First, for any given selection on the scatterplots, they can check a histogram showing the distribution of a concept's likelihood across the selected points. If the histogram is shifted to the right, it means the selection has a higher chance of belonging to a concept. 
Besides that, the users can assess the robustness of models in the Model Summary View (Figure~\ref{fig:visual_interface}(f)) where we present the evolution of the prototypes over the course of several refinements. Once we create a new version of a labeled subset for a specific concept, we train a new classification model to be part of the prototype and evaluate old versions of the prototype's classification model to assess the change in prediction over time. At some point, the user can come to a conclusion that labeling more points has no significant impact on the classification model and then stop the process.

\begin{figure*}[t!]
    \centering
    \includegraphics[width=1.0\linewidth]{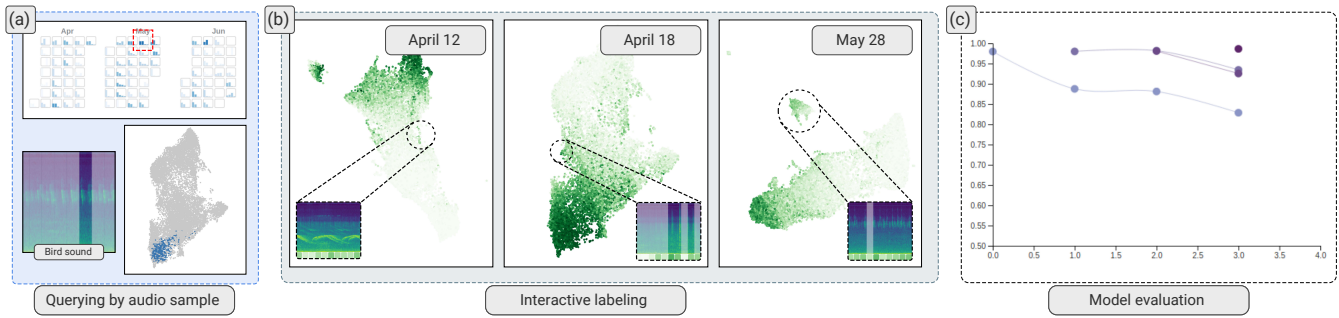}
    \caption{Interactive monitoring of the training process and refining the model. (a) We run a query using our sample birds' sound and analyze the clusters; (b) Investigating the clusters on different days to detect and re-label false positive and false negative instances, and refine the model; (c) The model evaluation indicates that our prototype models are converging as we do further iterations of refinement.}
    \label{fig:birds1}
\end{figure*}

%%%%%%%%%%%%%%%%%%%%%%%%%%%%%%%%%%%%%%%%%%%%%
\subsection{Analysis flow}
%%%%%%%%%%%%%%%%%%%%%%%%%%%%%%%%%%%%%%%%%%%%%

The exploration process starts with the user querying the data set using any of the three approaches we propose: select a frame from the examples we provide in a Query View as input for the similarity query, upload their own audio snippet and select a frame from this audio snippet, or query using one of the created prototypes. For all three query approaches, the user is able to select the number of frames the query will retrieve.
Once the query is processed, the Calendar View is updated, showing the density of a given class, or concept, on each cell throughout the year (color), and its distribution within the day (bar chart). Next, the user can select a specific day and load all the available data for that day to further inspect the day's soundscape using the scatterplots in the Day View.
At this point, the user can select specific regions of the scatterplot and listen to the correspondent audio frames, reproject specific regions of the day scatterplot to focus on local structures, remove undesired clusters or steer the scatterplot based on the annotation of frames. Also, color the points by prototype probability or created annotations. These operations will help users in two tasks: assess the performance of the prototypes they are creating and find data points that should be labeled as any concept of interest.
Following that, it's possible to create different prototypes and refine existing ones based on new annotations the users are creating, either positive annotations or negative. Meanwhile, when prototypes are created and refined, the Model Summary gets updated, showing the change in prediction probability of the models and the set of representative frames of a given concept. 
When the user is confident about the prototype they are creating, they can reuse this prototype to query the entire data set and look for specific temporal patterns that a specific concept is happening. This analysis flow denotes the importance of having a user in the loop to evaluate the performance of the prototype models as \name allows for the creation of concepts that match the user perception of the city's soundscape, which can not be evaluated quantitatively.

%%%%%%%%%%%%%%%%%%%%%%%%%%%%%%%%%%%%%%%%%%%%%
\subsection{System implementation}
%%%%%%%%%%%%%%%%%%%%%%%%%%%%%%%%%%%%%%%%%%%%%

We decided to develop \name following a client-server architecture. We structured our application following microservices guidelines to ensure that we could effortlessly add new features to the tool and scale its deployment to make it available for the general public. % (Figure~\ref{fig:architecture} presents an overview).
The storage component keeps audio recordings and their embeddings located in different folders following the same naming convention for faster localization. Each audio file is also associated with a set of metadata attributes with temporal and spatial information (time of the recording and location of the sensor) that is kept in a separate database.
The core of our application is composed of several microservices.
The data server is responsible to serve audio files and spectrogram images. 
The web server provides users with a bundle of our Angular web application.
The user server stores annotations on RocksDB~\cite{Doraiswamy:2018:IVE:3183713.3193559}.
The most complex services of our system are the ML server and the ANN server. The first is responsible for all machine-learning-related operations, such as multidimensional projections, hierarchical clustering, and model training. Following \textbf{R6}, the operations are processed using GPUs through RAPIDS libraries~\cite{rapids}. CPU-based libraries would not be able to handle such data-intensive operations required by \name. The ANN server is responsible for computing similarity queries based on the euclidean distance between frames.

%% file: sections/7-cases.tex
\section{Case Studies}
In this section, we demonstrate the application of \name through two case studies using data from the SONYC sensors. In doing so, we highlight how the requirements listed in Section~\ref{sec:requirements} are met in different tasks.
The first case study explores how \name can facilitate the interactive labeling and exploration of data for investigating out-of-hours construction noise, a pressing issue facing many large cities.
The second one highlights another capability of \name to facilitate searching for mixture of sounds to explore the impact of anthropogenic noises such as siren on bird songs. 
These case studies can be of interest to various stakeholders, from the general public and advocacy groups to government agencies, such as the Dept. of Environmental Protection.

\begin{figure*}[t!]
    \centering
    \includegraphics[width=1.0\linewidth]{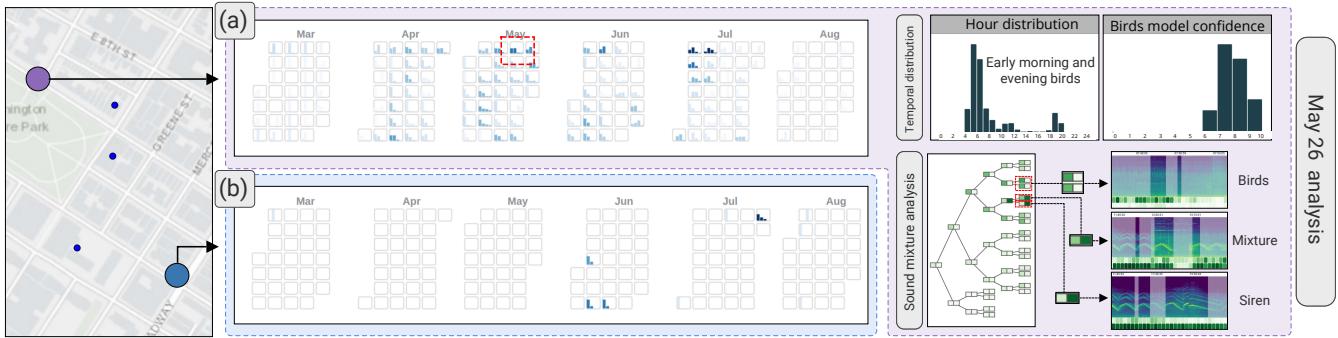}
    \caption{Looking for bird songs in two different Manhattan locations: (a) Edge of Washington Square Park with high concentration of bird songs and (b) a street corner on Broadway with very few instances of bird songs since we do not have trees for birds to nest.}
    \label{fig:birds2}
\end{figure*}

\subsection{After-hour construction noise}

Construction noise is one of the primary sources of noise-related complaints in NYC. As the city grows, new structures are built, old ones get renovated, and economic pressures and deadlines lead developers to request the city for permits allowing them to perform construction outside the regular workday hours (i.e., 8~AM to 5~PM). In the past few years, this has been a major source of dispute between NYC residents and developers~\cite{mays_why_2019}, and this problem is increasingly getting worse. In 2018, NYC's Department of Buildings issued around 67,000 after-hour permits, more than double the number of permits issued in 2012. Although developers must follow strict noise guidelines during after-hour constructions, the increase in the number of complaints related to these types of disturbances indicates otherwise.
Even though the city constantly issues noise construction fines through manual inspections, the after-hour nature of these noises makes it especially hard to monitor them. This is a significant problem that needs to be addressed by cities and their different departments, with severe political, social, and economic ramifications.

In this study, we use the SONYC network to understand the impact of construction-related noises on the soundscape of NYC. Our first goal is to assess if these noises were captured by our sensors, to facilitate noise code enforcement activities. Secondly, we would like to use examples that we found during our initial exploration to build a prototype capable of pointing us to specific days and times where after-hour construction work might have happened. 
We start by querying our data set for similar audio snippets using one of the examples provided in the system containing the recording of a powered saw~\textbf{(R1)}. Using the Calendar View, we can quickly observe a day containing most of the similar audio excerpts according to our ANN model~(Figure~\ref{fig:teaser}(a, top)).
We select that day, and \name generates a UMAP projection of all the audio frames within that day~(Figure~\ref{fig:teaser}(a, bottom)). After a quick inspection of the projection scatterplot, we can notice a set of distinctive clusters (highlighted in red). Using the tool's interactions, we start by selecting the one cluster containing most of the points retrieved by the initial similarity query.
By listening to a few recordings, we can notice that the points belonging to this cluster are perceptually similar to a powered saw, very common on construction sites \textbf{(R2)}. We also notice that most of these audio snippets were recorded around 8 AM, as the hour distribution chart shows us.
Figure~\ref{fig:teaser}(b,c) highlights the recordings that happened around 8 AM, and it's possible to again see different clusters.
After listening to recordings from each cluster, we noticed that each one of them represents different sounds (powered saw, drilling machine, engine).
At this point, we can leverage \name's feature that allows us to create models on the fly and decide whether to include certain sounds in our prototype~\textbf{(R3)}.
Once we label recordings from that specific day, we generate two construction prototypes (with and without large engine noise).
We can now use them to guide the exploration through different days of the year.
This step allows us to speed up the search for similar sounds, without the need to listen to \emph{hours and hours} of soundscape audio files.
Also, during this guided exploration, we can adjust the prototype by labeling more points, either as negative or positive labels, as we assess the model's performance by listening to the recordings. This interactive process is highlighted in Figure~\ref{fig:teaser}(b,c), for two different models.

After refining our models once, we listened to the representative snippets of our prototypes and used them to look for unusual events.
The calendar heatmaps show the results of the prototype queries (Figure~\ref{fig:teaser}(d,e)) where we can spot two interesting events.
In February, we noticed that during two days, construction work happened during the night (Figure~\ref{fig:teaser}(d)). And that, during many days in October, the same engine noise started at 11~PM and lasted for approximately 30 minutes (Figure~\ref{fig:teaser}(e)). 

To further validate this finding, we used citizen complaints filled through NYC's 311 non-emergency service phone number. Interestingly, there were actually a series of complaints reported on those two specific days of February.
The ability to intuitively create prototypes based on audio files listened in the exploratory process sets \name apart. Findings such as these not only highlight the usefulness of a \emph{passive} network of sensors (as opposed to \emph{active} sensors deployed in inspection visits), but also the usefulness of distinguishing different noises emitted from construction sites. Previous approaches, like Noise Profiler~\cite{miranda_time_2018}, focus on the SPL characteristics, a useful but crude measurement of noise. By enabling the exploration of specific types of noise, \name can 1)~provide a clearer picture of the soundscape near a construction site, 2)~facilitate monitoring tasks carried out by enforcement agencies, and 3)~validate the accuracy of 311 complaints.

\subsection{Birds in New York City}
The impacts of urban noise, air pollution, and the built environment on residents and migrating birds have been extensively studied~\cite{seress2015habitat}. There is a strand of research that specifically analyze birdsong to discover if exposure to loud urban noise can lead to significant changes in their song traits and the time and frequency of their chorus, specifically since birds use different sounds to communicate, mate, and defend breeding territories and rely on the vocal communication to sustain their lives~\cite{mendes2011bird, slabbekoorn2013songs}.  
One of the main challenges in the majority of bioacoustics and avian behavior studies is the costly and time-consuming nature of working with audio data, which limits the duration and geographical extent of the research. The application of machine learning in bird song classification is not new~\cite{mcilraith1997bird}, but most of the developed models are trained using specific sets of data, limiting the user to a pre-defined set of labels, with no control over what the model perceives as bird songs. This is specifically important in bird song studies since the model can classify some sounds, such as whistling, as bird sounds and discard some bird songs which are very different from what it was trained on~\cite{XIE201974}. 

In this case study, we demonstrate how \name can facilitate such studies by providing a robust and easy to use solution where the user can search for specific sounds among hundreds of hours of recordings, refine the results if needed to reach the confidence level of interest, monitor the frequency and changes in the song traits, and investigate the impact of anthropogenic noises on birds. 
Sitting on the Atlantic Flyway, NYC offers great resting grounds for birds traveling along the north-south migratory route in the Americas~\cite{day2015field}. We choose Washington Square Park, a popular local park situated in a dense and busy neighborhood of the Manhattan borough, with the natural environment for birds to nest as well as the attributes representing a crowded and noisy urban environment~\cite{wspbirds}. 

\begin{figure}[b!]
    \centering
    \includegraphics[width=1.0\linewidth]{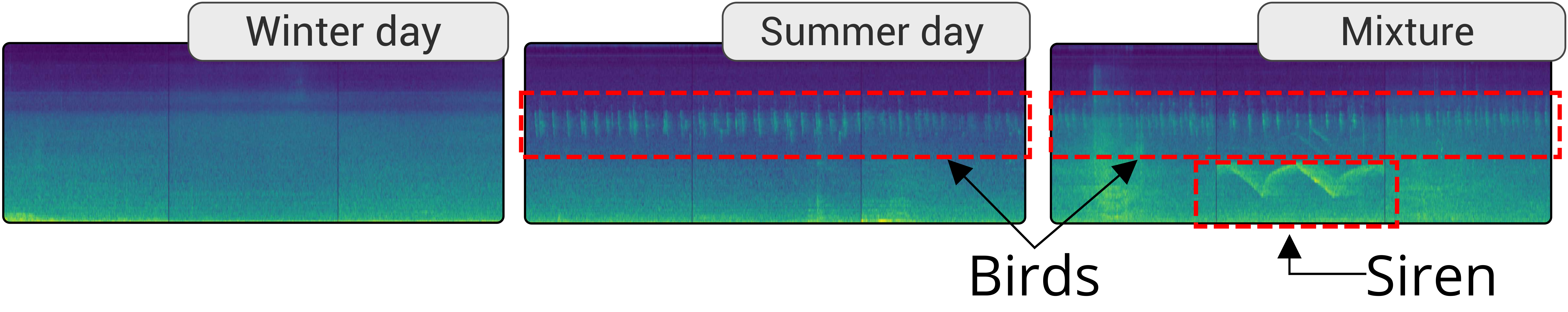}
      \caption{Spectrogram showing a winter day with no bird song, a summer day with birds' singing and the selected day in summer when birds dawn chorus continued despite loud siren.}
    \label{fig:birds3}
\end{figure}

The first step is to build our bird representation model. We start our exploration by using one of the bird song examples provided in the query view. Next, we select a day with high density of similar bird sounds. As shown in Figure~\ref{fig:birds1}(a), we generate a UMAP projection of our selected day on the Day View and see that majority of the bird songs are clustered on the bottom region of the projection (blue points). Next, we create our first representation model of bird songs to speed up our search across different days~\textbf{(R1)}. We can find false positives and false negative examples throughout this process, fix those and refine our prototype. For instance, we found out that on April 18th, the model assigned a high likelihood to a small cluster of points (Figure~\ref{fig:birds1}(b)). We investigate this cluster closely and realize they are not bird songs, so we re-label these points, refine our model, make a new prediction with updated weights, and run this process iteratively until the model reaches a robust state~\textbf{(R2)}. 
In the Model Summary View (Figure~\ref{fig:birds1}(c)), we can see that our new prototypes are converging: Our first model had the worst performance, and as we continued refining, the difference between the prediction probabilities of the labeled birds' data set get smaller after each iteration~\textbf{(R3)}.

Using our refined model, we run a new query to explore the distribution and patterns of bird songs near Washington Square Park over the course of one year. The retrieved results clearly show two levels of seasonal patterns: a daily pattern with peaks in the mornings and afternoons corresponding to the dawn and dusk chorus times, and another pattern with peaks during spring to early summer, when songbirds usually migrate, as illustrated in Figure~\ref{fig:birds2}(a). This signifies the robust performance of the model in classifying birds.
We also look at the corner of Broadway and Waverly Pl., where we have no trees on both sides of the street, to see if we can find similar patterns there. As Figure~\ref{fig:birds2}(b) shows, we have very few instances of bird songs in that location throughout 2017.

One useful aspect of \name is the ability to analyze sound mixtures. To investigate how the siren sound can impact or even halt the birds' chorus, we use \name to query for dawn chorus times (6-11 AM) where siren was also present.
This allows us to discover whether loud sirens can halt birds' dawn chorus or whether birds in noisy urban areas like Manhattan local parks are adapted to the level of noise~\cite{nemeth2010birds, nemeth2013bird}.
We can use the Mixture Explorer to differentiate between these two sounds, as illustrated by Figure~\ref{fig:birds2}(a, bottom right). Notice that nodes containing bird songs, siren, or mixture of both are clearly distinguishable with our visual encodings~\textbf{(R4)}. Drilling down to this specific example (Figure~\ref{fig:birds3}), we can see that the birds continue singing despite the loud siren~\textbf{(R5)}.
This analysis can create a ground for further research by bioacousticians and researchers in this field to investigate whether this pattern is more prevalent in birds of specific species or whether we can find incidents of ambient noise halting birds singing.
\name helped us to iteratively refine our model, track the sounds of interest and search for a combination of sounds across a large data set, detect the pattern and drill down to the exact moments to listen and investigate more.

%% file: sections/8-conclusion.tex
\section{Discussion and Conclusion}
We have presented \name, a novel interactive system for seamlessly exploring large audio data sets, based on user-generated concepts. Leveraging machine learning techniques, \name supports labeling and analysis at scale, while our multilevel visualization approach enables the inspection of temporal patterns at varying levels of granularity. By enabling users to interactively label data based on their knowledge, \name can be used to augment self-supervised methods that might not account for audio complexity. We illustrate its potential through data collected by the SONYC project. However, \name can be applied to other longitudinal spatiotemporal acoustic data (e.g., bioacustics~\cite{underwaterdata, andrew_farnsworth_2021_5856260}), and to support this we made the tool available on GitHub. We hope this will encourage researchers to use it in many different contexts and further develop the code base.

\myparagraph{Limitations.}
While we define interactivity based on benchmarks for querying large data ~\cite{battle2020database}, we also identify three potential bottlenecks: similarity search, model training, and projection generation. \name  responds to similarity queries by returning up to 10,000 points in less than one second (for the examples provided as initial query seeds~\cite{faiss}). However, a one-time preprocessing computation is required to generate indices. This takes on average one hour per sensor/year and needs 9~GB of memory space (for sensors with low rates of missing data). GPU implementations~\cite{rapids, rapidsbenchmark} achieve response times of under one second when loading Day View selections and for inference of created concept models. Deploying \name to handle data from alternate sensor networks requires sufficient memory space to handle query indices, GPU capabilities to train models, and connectivity to support client-server architectures.

\myparagraph{Expert feedback.}
Analyzing large collections of audio data is a challenging task, in which views into the data can be limited. The number of classes classifiers detect may be small,  not matched to the task at hand, or too coarse-grained. Deep audio embeddings help to distill the semantics of audio to a smaller number of dimensions, but they are still very opaque and not easily interpretable. In addition, translation between modalities (e.g., using visual tools to explore audio data) is also highly challenging, and yet we know that it can be very effective. Our collaborators highlighted that \name helps overcome these challenges by enabling interactive  exploration, labeling, clustering, and reprojection of collections of audio data; and supports insights into models, labeled data, and previously unseen patterns within unlabeled data.

\myparagraph{Future work.} We plan to investigate whether \name can accurately and efficiently represent concepts matching the user's mental model of their data. To investigate this we plan to conduct a large-scale user study with machine learning and audio researchers. While previous research~\cite{cartwright2017seeing} shows that spectrogram visualizations lead to high annotation accuracy at low time and labor costs, further investigation is also needed to explore additional visualization metaphors (e.g., to summarize longer periods of audio recordings). We will also explore how the analyses supported by systems such as \name can useful to public officials and community representatives.

\myparagraph{Conclusion.} \name is an interactive visual analytics tool for gaining insight into large collections of audio data, which we have demonstrated through use cases that characterize the acoustic environment of NYC. We believe that \name offers an important step in moving beyond simple metrics, such as SPL, and will be of value to researchers in human-centered machine learning, acoustics, and urban science.

%% file: sections/acknoledgements.tex
\section*{Acknowledgements}

We would like to thank our colleagues at CUSP (NYU) for their feedback during the development of this work.
This research has been supported by NSF awards CNS-1229185, CCF-1533564, CNS-1544753, CNS-1730396, CNS-1828576, CNS-1626098; CNPq grant 305974/2018-1; FAPERJ grants E-26/202.915/2019, E-26/211.134/2019. 

% Any opinions, findings, and conclusions or recommendations expressed in this material are those of the authors and do not necessarily reflect the views of NSF.

%% file: main.bbl
\newcommand{\etalchar}[1]{$^{#1}$}
\begin{thebibliography}{\uppercase{VKDSVK14}}

\bibitem[AA08]{andrienko_spatio-temporal_2008}
\textsc{Andrienko G., Andrienko N.}:
\newblock Spatio-temporal aggregation for visual analysis of movements.
\newblock In \emph{2008 {IEEE} Symposium on Visual Analytics Science and
  Technology} (2008), IEEE, pp.~51--58.

\bibitem[AVT16]{aytar_soundnet_2016}
\textsc{Aytar Y., Vondrick C., Torralba A.}:
\newblock Soundnet: {Learning} sound representations from unlabeled video.
\newblock \emph{arXiv preprint ID:1610.09001} (2016).

\bibitem[AZ17]{arandjelovic_look_2017}
\textsc{Arandjelovic R., Zisserman A.}:
\newblock Look, listen and learn.
\newblock In \emph{Proceedings of the {IEEE} {International} {Conference} on
  {Computer} {Vision}} (2017), pp.~609--617.

\bibitem[BB20]{sonycnyt2020}
\textsc{Bui Q., Badger E.}:
\newblock {The Coronavirus Quieted City Noise. Listen to What’s Left.}
\newblock \emph{The New York Times} (May 2020).
\newblock URL:
  \url{https://www.nytimes.com/interactive/2020/05/22/upshot/coronavirus-quiet-city-noise.html}.

\bibitem[BEA{\etalchar{*}}20]{battle2020database}
\textsc{Battle L., Eichmann P., Angelini M., Catarci T., Santucci G., Zheng Y.,
  Binnig C., Fekete J.-D., Moritz D.}:
\newblock Database benchmarking for supporting real-time interactive querying
  of large data.
\newblock In \emph{Proceedings of the 2020 International Conference on
  Management of Data} (2020), SIGMOD '20, ACM, pp.~1571--1587.

\bibitem[BH10]{bronzaft_noise_2010}
\textsc{Bronzaft A.~L., Hagler L.}:
\newblock Noise: {The} invisible pollutant that cannot be ignored.
\newblock In \emph{Emerging {Environmental} {Technologies}, {Volume} {II}}.
  Springer, 2010, pp.~75--96.

\bibitem[BHZ{\etalchar{*}}17]{bernard2017comparing}
\textsc{Bernard J., Hutter M., Zeppelzauer M., Fellner D., Sedlmair M.}:
\newblock Comparing visual-interactive labeling with active learning: An
  experimental study.
\newblock \emph{IEEE Transactions on Visualization and Computer Graphics 24}, 1
  (2017), 298--308.

\bibitem[Bro07]{bronzaft_neighborhood_2007}
\textsc{Bronzaft A.}:
\newblock Neighborhood noise and its consequences.
\newblock \emph{Survey Research Unit, School of Public Affairs, Baruch College,
  New York} (2007).

\bibitem[Bro10]{brown_soundscapes_2010}
\textsc{Brown A.~L.}:
\newblock Soundscapes and environmental noise management.
\newblock \emph{Noise Control Engineering Journal 58}, 5 (2010), 493--500.

\bibitem[Bro12]{brown_review_2012}
\textsc{Brown A.~L.}:
\newblock A review of progress in soundscapes and an approach to soundscape
  planning.
\newblock \emph{International Journal of Acoustics and Vibration 17}, 2 (2012),
  73--81.

\bibitem[BSN{\etalchar{*}}19]{bello_sonyc_2019}
\textsc{Bello J.~P., Silva C., Nov O., Dubois R.~L., Arora A., Salamon J.,
  Mydlarz C., Doraiswamy H.}:
\newblock Sonyc: {A} system for monitoring, analyzing, and mitigating urban
  noise pollution.
\newblock \emph{Communications of the ACM 62}, 2 (2019), 68--77.

\bibitem[CCSB19]{cartwright_tricycle_2019}
\textsc{Cartwright M., Cramer J., Salamon J., Bello J.~P.}:
\newblock {TriCycle}: {Audio} representation learning from sensor network data
  using self-supervision.
\newblock In \emph{2019 {IEEE} {Workshop} on {Applications} of {Signal}
  {Processing} to {Audio} and {Acoustics} ({WASPAA})} (2019), IEEE,
  pp.~278--282.

\bibitem[CDDF16]{chirigati_data_2016}
\textsc{Chirigati F., Doraiswamy H., Damoulas T., Freire J.}:
\newblock Data polygamy: the many-many relationships among urban
  spatio-temporal data sets.
\newblock In \emph{Procedings of the 2016 {International} {Conference} on
  {Management} of {Data}} (2016), pp.~1011--1025.

\bibitem[CSS{\etalchar{*}}17]{cartwright2017seeing}
\textsc{Cartwright M., Seals A., Salamon J., Williams A., Mikloska S.,
  MacConnell D., Law E., Bello J.~P., Nov O.}:
\newblock Seeing sound: Investigating the effects of visualizations and
  complexity on crowdsourced audio annotations.
\newblock \emph{Proceedings of the ACM on Human-Computer Interaction 1}, CSCW
  (2017), 1--21.

\bibitem[CWSB19]{cramer_look_2019}
\textsc{Cramer J., Wu H.-H., Salamon J., Bello J.~P.}:
\newblock Look, listen, and learn more: {Design} choices for deep audio
  embeddings.
\newblock In \emph{2019 {IEEE} {International} {Conference} on {Acoustics},
  {Speech} and {Signal} {Processing} ({ICASSP})} (2019), IEEE, pp.~3852--3856.

\bibitem[DAB{\etalchar{*}}13]{davies_perception_2013}
\textsc{Davies W.~J., Adams M.~D., Bruce N.~S., Cain R., Carlyle A., Cusack P.,
  Hall D.~A., Hume K.~I., Irwin A., Jennings P.}:
\newblock Perception of soundscapes: {An} interdisciplinary approach.
\newblock \emph{Applied acoustics 74}, 2 (2013), 224--231.

\bibitem[DBC{\etalchar{*}}17]{dema2017collaborative}
\textsc{Dema T., Brereton M., Cappadonna J.~L., Roe P., Truskinger A., Zhang
  J.}:
\newblock Collaborative exploration and sensemaking of big environmental sound
  data.
\newblock \emph{Computer Supported Cooperative Work 26}, 4–6 (2017),
  693–731.

\bibitem[DFL{\etalchar{*}}18]{8474495}
\textsc{Doraiswamy H., Freire J., Lage M., Miranda F., Silva C.}:
\newblock Spatio-temporal urban data analysis: A visual analytics perspective.
\newblock \emph{IEEE Computer Graphics and Applications 38}, 5 (2018), 26--35.

\bibitem[dPVCR15]{de_paiva_vianna_noise_2015}
\textsc{de~Paiva~Vianna K.~M., Cardoso M. R.~A., Rodrigues R. M.~C.}:
\newblock Noise pollution and annoyance: {An} urban soundscapes study.
\newblock \emph{Noise \& Health 17}, 76 (2015), 125.

\bibitem[DR15]{day2015field}
\textsc{Day L., Riepe D.}:
\newblock \emph{Field Guide to the Neighborhood Birds of New York City}.
\newblock JHU Press, 2015.

\bibitem[DTZM{\etalchar{*}}18]{Doraiswamy:2018:IVE:3183713.3193559}
\textsc{Doraiswamy H., Tzirita~Zacharatou E., Miranda F., Lage M., Ailamaki A.,
  Silva C.~T., Freire J.}:
\newblock Interactive visual exploration of spatio-temporal urban data sets
  using urbane.
\newblock In \emph{Proceedings of the 2018 International Conference on
  Management of Data} (2018), SIGMOD '18, ACM, pp.~1693--1696.

\bibitem[DWX{\etalchar{*}}21]{deng2021compass}
\textsc{Deng Z., Weng D., Xie X., Bao J., Zheng Y., Xu M., Chen W., Wu Y.}:
\newblock Compass: Towards better causal analysis of urban time series.
\newblock \emph{IEEE Transactions on Visualization and Computer Graphics 28}, 1
  (2021), 1051--1061.

\bibitem[DZD{\etalchar{*}}10]{dratva_impact_2010}
\textsc{Dratva J., Zemp E., Dietrich D.~F., Bridevaux P.-O., Rochat T.,
  Schindler C., Gerbase M.~W.}:
\newblock Impact of road traffic noise annoyance on health-related quality of
  life: {Results} from a population-based study.
\newblock \emph{Quality of Life Research 19}, 1 (2010), 37--46.

\bibitem[Fai]{faiss}
{Faiss}.
\newblock URL:~\url{https://faiss.ai/}.

\bibitem[FFP{\etalchar{*}}20]{fonseca_fsd50k_2020}
\textsc{Fonseca E., Favory X., Pons J., Font F., Serra X.}:
\newblock {FSD50k}: an open dataset of human-labeled sound events.
\newblock \emph{arXiv preprint ID:2010.00475} (2020).

\bibitem[FKL{\etalchar{*}}21]{andrew_farnsworth_2021_5856260}
\textsc{Farnsworth A., Kelling S., Lostanlen V., Salamon J., Cramer A., Bello
  J.~P.}:
\newblock {BirdVox-296h: a large-scale dataset for detection and classification
  of flight calls}, Dec. 2021.

\bibitem[FLD{\etalchar{*}}15]{ferreira_urbane_2015}
\textsc{Ferreira N., Lage M., Doraiswamy H., Vo H., Wilson L., Werner H., Park
  M., Silva C.}:
\newblock Urbane: {A} {3D} framework to support data driven decision making in
  urban development.
\newblock In \emph{2015 {IEEE} Conference on Visual Analytics Science and
  Technology ({VAST})} (2015), IEEE, pp.~97--104.

\bibitem[FPV{\etalchar{*}}13]{ferreira_visual_2013}
\textsc{Ferreira N., Poco J., Vo H.~T., Freire J., Silva C.~T.}:
\newblock Visual exploration of big spatio-temporal urban data: {A} study of
  new york city taxi trips.
\newblock \emph{IEEE Transactions on Visualization and Computer Graphics 19},
  12 (2013), 2149--2158.

\bibitem[GCA06]{guite_impact_2006}
\textsc{Guite H.~F., Clark C., Ackrill G.}:
\newblock The impact of the physical and urban environment on mental
  well-being.
\newblock \emph{Public Health 120}, 12 (2006), 1117--1126.

\bibitem[GCKT21]{grollmisch_analyzing_2021}
\textsc{Grollmisch S., Cano E., Kehling C., Taenzer M.}:
\newblock Analyzing the {Potential} of {Pre}-{Trained} {Embeddings} for {Audio}
  {Classification} {Tasks}.
\newblock In \emph{2020 28th {European} {Signal} {Processing} {Conference}
  ({EUSIPCO})} (2021), IEEE, pp.~790--794.

\bibitem[GEF{\etalchar{*}}17]{gemmeke_audio_2017}
\textsc{Gemmeke J.~F., Ellis D.~P., Freedman D., Jansen A., Lawrence W., Moore
  R.~C., Plakal M., Ritter M.}:
\newblock Audio set: {An} ontology and human-labeled dataset for audio events.
\newblock In \emph{2017 {IEEE} {International} {Conference} on {Acoustics},
  {Speech} and {Signal} {Processing} ({ICASSP})} (2017), IEEE, pp.~776--780.

\bibitem[Gua03]{guastavino_etude_2003}
\textsc{Guastavino C.}:
\newblock \emph{Etude sémantique et acoustique de la perception des basses
  fréquences dans l'environnement sonore urbain}.
\newblock {PhD} {Thesis}, Paris 6, 2003.

\bibitem[HCE{\etalchar{*}}17]{hershey_cnn_2017}
\textsc{Hershey S., Chaudhuri S., Ellis D.~P., Gemmeke J.~F., Jansen A., Moore
  R.~C., Plakal M., Platt D., Saurous R.~A., Seybold B., {others}}:
\newblock {CNN} architectures for large-scale audio classification.
\newblock In \emph{2017 {IEEE} International Conference on Acoustics, Speech
  and Signal Processing ({ICASSP})} (2017), IEEE, pp.~131--135.

\bibitem[HDVT{\etalchar{*}}08]{haralabidis_acute_2008}
\textsc{Haralabidis A.~S., Dimakopoulou K., Vigna-Taglianti F., Giampaolo M.,
  Borgini A., Dudley M.-L., Pershagen G., Bluhm G., Houthuijs D., Babisch W.}:
\newblock Acute effects of night-time noise exposure on blood pressure in
  populations living near airports.
\newblock \emph{European Heart Journal 29}, 5 (2008), 658--664.

\bibitem[HSN14]{hammer_environmental_2014}
\textsc{Hammer M.~S., Swinburn T.~K., Neitzel R.~L.}:
\newblock Environmental noise pollution in the {United} {States}: developing an
  effective public health response.
\newblock \emph{Environmental Health Perspectives 122}, 2 (2014), 115--119.

\bibitem[IYT{\etalchar{*}}14]{itoh_visual_2014}
\textsc{Itoh M., Yokoyama D., Toyoda M., Tomita Y., Kawamura S., Kitsuregawa
  M.}:
\newblock Visual fusion of mega-city big data: an application to traffic and
  tweets data analysis of metro passengers.
\newblock In \emph{2014 {IEEE} {International} {Conference} on {Big} {Data}
  ({Big} {Data})} (2014), IEEE, pp.~431--440.

\bibitem[JCC{\etalchar{*}}11]{joia_local_2011}
\textsc{Joia P., Coimbra D., Cuminato J.~A., Paulovich F.~V., Nonato L.~G.}:
\newblock Local affine multidimensional projection.
\newblock \emph{IEEE Transactions on Visualization and Computer Graphics 17},
  12 (2011), 2563--2571.

\bibitem[JPP{\etalchar{*}}18]{jansen_unsupervised_2018}
\textsc{Jansen A., Plakal M., Pandya R., Ellis D.~P., Hershey S., Liu J., Moore
  R.~C., Saurous R.~A.}:
\newblock Unsupervised learning of semantic audio representations.
\newblock In \emph{2018 {IEEE} International Conference on Acoustics, Speech
  and Signal Processing ({ICASSP})} (2018), IEEE, pp.~126--130.

\bibitem[JVHKL01]{johnston_using_2001}
\textsc{Johnston K., Ver~Hoef J.~M., Krivoruchko K., Lucas N.}:
\newblock \emph{Using {ArcGIS} geostatistical analyst}, vol.~380.
\newblock Esri Redlands, 2001.

\bibitem[KKF18]{kumar_knowledge_2018}
\textsc{Kumar A., Khadkevich M., Fügen C.}:
\newblock Knowledge transfer from weakly labeled audio using convolutional
  neural network for sound events and scenes.
\newblock In \emph{2018 {IEEE} {International} {Conference} on {Acoustics},
  {Speech} and {Signal} {Processing} ({ICASSP})} (2018), IEEE, pp.~326--330.

\bibitem[LGG{\etalchar{*}}17]{lin2017rclens}
\textsc{Lin H., Gao S., Gotz D., Du F., He J., Cao N.}:
\newblock Rclens: Interactive rare category exploration and identification.
\newblock \emph{IEEE Transactions on Visualization and Computer Graphics 24}, 7
  (2017), 2223--2237.

\bibitem[LGTR15]{lenormand_human_2015}
\textsc{Lenormand M., Gonçalves B., Tugores A., Ramasco J.~J.}:
\newblock Human diffusion and city influence.
\newblock \emph{Journal of The Royal Society Interface 12}, 109 (2015),
  20150473.

\bibitem[LH14]{liu_effects_2014}
\textsc{Liu Z., Heer J.}:
\newblock The effects of interactive latency on exploratory visual analysis.
\newblock \emph{IEEE Transactions on Visualization and Computer Graphics 20},
  12 (2014), 2122--2131.

\bibitem[LJLH19]{liu_latent_2019}
\textsc{Liu Y., Jun E., Li Q., Heer J.}:
\newblock Latent space cartography: Visual analysis of vector space embeddings.
\newblock \emph{Computer Graphics Forum 38}, 3 (2019), 67--78.

\bibitem[LYC10]{liao2010anomaly}
\textsc{Liao Z., Yu Y., Chen B.}:
\newblock Anomaly detection in gps data based on visual analytics.
\newblock In \emph{2010 IEEE Symposium on Visual Analytics Science and
  Technology} (2010), IEEE, pp.~51--58.

\bibitem[May19]{mays_why_2019}
\textsc{Mays J.~C.}:
\newblock Why {Construction} {Noise} {Is} {Keeping} {You} {Up} at 3 {A}.{M}.
\newblock \emph{The New York Times} (Sept. 2019).
\newblock URL:
  \url{https://www.nytimes.com/2019/09/27/nyregion/noise-construction-sleep-nyc.html}.

\bibitem[MC97]{mcilraith1997bird}
\textsc{McIlraith A., Card H.}:
\newblock Bird song identification using artificial neural networks and
  statistical analysis.
\newblock In \emph{CCECE'97. Canadian Conference on Electrical and Computer
  Engineering. Engineering Innovation: Voyage of Discovery. Conference
  Proceedings} (1997), vol.~1, IEEE, pp.~63--66.

\bibitem[MCRP11]{mendes2011bird}
\textsc{Mendes S., Colino-Rabanal V.~J., Peris S.~J.}:
\newblock Bird song variations along an urban gradient: The case of the
  european blackbird (turdus merula).
\newblock \emph{Landscape and Urban Planning 99}, 1 (2011), 51--57.

\bibitem[MDL{\etalchar{*}}17]{7539380}
\textsc{Miranda F., Doraiswamy H., Lage M., Zhao K., Gon\c{c}alves B., Wilson
  L., Hsieh M., Silva C.~T.}:
\newblock {Urban Pulse}: {C}apturing the rhythm of cities.
\newblock \emph{IEEE Transactions on Visualization and Computer Graphics 23}, 1
  (2017), 791--800.

\bibitem[MDL{\etalchar{*}}19]{8283638}
\textsc{Miranda F., Doraiswamy H., Lage M., Wilson L., Hsieh M., Silva C.~T.}:
\newblock {Shadow Accrual Maps}: Efficient accumulation of city-scale shadows
  over time.
\newblock \emph{IEEE Transactions on Visualization and Computer Graphics 25}, 3
  (2019), 1559--1574.

\bibitem[MHL{\etalchar{*}}20]{10.1145/3313831.3376399}
\textsc{Miranda F., Hosseini M., Lage M., Doraiswamy H., Dove G., Silva C.~T.}:
\newblock {Urban Mosaic}: Visual exploration of streetscapes using large-scale
  image data.
\newblock In \emph{Proceedings of the 2020 CHI Conference on Human Factors in
  Computing Systems} (2020), CHI ’20, ACM, p.~1–15.

\bibitem[MHM18]{mcinnes_umap_2018}
\textsc{McInnes L., Healy J., Melville J.}:
\newblock Umap: {Uniform} manifold approximation and projection for dimension
  reduction.
\newblock \emph{arXiv preprint ID:1802.03426} (2018).

\bibitem[MLD{\etalchar{*}}18]{miranda_time_2018}
\textsc{Miranda F., Lage M., Doraiswamy H., Mydlarz C., Salamon J., Lockerman
  Y., Freire J., Silva C.~T.}:
\newblock {Time Lattice}: {A} data structure for the interactive visual
  analysis of large time series.
\newblock \emph{Computer Graphics Forum 37}, 3 (2018), 23--35.

\bibitem[MME{\etalchar{*}}12]{malik2012correlative}
\textsc{Malik A., Maciejewski R., Elmqvist N., Jang Y., Ebert D.~S., Huang W.}:
\newblock A correlative analysis process in a visual analytics environment.
\newblock In \emph{2012 IEEE Conference on Visual Analytics Science and
  Technology (VAST)} (2012), IEEE, pp.~33--42.

\bibitem[MMW]{underwaterdata}
\textsc{Miller B.~S., Milnes M., Whiteside S.}:
\newblock Long-term underwater acoustic recordings 2013-2019.
\newblock
  URL:~\url{https://researchdata.edu.au/long-term-underwater-2013-2019/967510}.

\bibitem[Muz02]{muzet_need_2002}
\textsc{Muzet A.}:
\newblock The need for a specific noise measurement for population exposed to
  aircraft noise during night-time.
\newblock \emph{Noise and Health 4}, 15 (2002), 61.

\bibitem[NB10]{nemeth2010birds}
\textsc{Nemeth E., Brumm H.}:
\newblock Birds and anthropogenic noise: are urban songs adaptive?
\newblock \emph{The American Naturalist 176}, 4 (2010), 465--475.

\bibitem[NGM{\etalchar{*}}12]{neitzel_exposures_2012}
\textsc{Neitzel R.~L., Gershon R.~R., McAlexander T.~P., Magda L.~A., Pearson
  J.~M.}:
\newblock Exposures to transit and other sources of noise among {New} {York}
  {City} residents.
\newblock \emph{Environmental science \& technology 46}, 1 (2012), 500--508.

\bibitem[NKLM20]{nadj2020power}
\textsc{Nadj M., Knaeble M., Li M.~X., Maedche A.}:
\newblock Power to the oracle? {D}esign principles for interactive labeling
  systems in machine learning.
\newblock \emph{KI-K{\"u}nstliche Intelligenz 34}, 2 (2020), 131--142.

\bibitem[NPZ{\etalchar{*}}13]{nemeth2013bird}
\textsc{Nemeth E., Pieretti N., Zollinger S.~A., Geberzahn N., Partecke J.,
  Miranda A.~C., Brumm H.}:
\newblock Bird song and anthropogenic noise: vocal constraints may explain why
  birds sing higher-frequency songs in cities.
\newblock \emph{Proceedings of the Royal Society B: Biological Sciences 280},
  1754 (2013), 20122798.

\bibitem[NSL{\etalchar{*}}12]{noulas_tale_2012}
\textsc{Noulas A., Scellato S., Lambiotte R., Pontil M., Mascolo C.}:
\newblock A tale of many cities: universal patterns in human urban mobility.
\newblock \emph{PloS one 7}, 5 (2012), e37027.

\bibitem[Org11]{organization_burden_2011}
\textsc{Organization W.~H.}:
\newblock \emph{Burden of disease from environmental noise: {Quantification} of
  healthy life years lost in {Europe}}.
\newblock World Health Organization. Regional Office for Europe, 2011.

\bibitem[OSS{\etalchar{*}}16]{ortner_vis--ware_2016}
\textsc{Ortner T., Sorger J., Steinlechner H., Hesina G., Piringer H., Gröller
  E.}:
\newblock Vis-a-ware: {Integrating} spatial and non-spatial visualization for
  visibility-aware urban planning.
\newblock \emph{IEEE Transactions on Visualization and Computer Graphics 23}, 2
  (2016), 1139--1151.

\bibitem[PDA09]{payne_research_2009}
\textsc{Payne S.~R., Davies W.~J., Adams M.~D.}:
\newblock \emph{Research into the practical and policy applications of
  soundscape concepts and techniques in urban areas}.
\newblock Tech. rep., University of Salford, 2009.

\bibitem[QS14]{quercia_mining_2014}
\textsc{Quercia D., Saez D.}:
\newblock Mining urban deprivation from foursquare: {Implicit} crowdsourcing of
  city land use.
\newblock \emph{IEEE Pervasive Computing 13}, 2 (2014), 30--36.

\bibitem[Rapa]{rapids}
{RAPIDS}.
\newblock URL:~\url{https://rapids.ai/start.html}.

\bibitem[Rapb]{rapidsbenchmark}
{RAPIDS Benchmark}.
\newblock
  URL:~\url{https://www.alcf.anl.gov/sites/default/files/2021-03/NVIDIA_RAPIDS_ANL.pdf}.

\bibitem[RD05]{raimbault_urban_2005}
\textsc{Raimbault M., Dubois D.}:
\newblock Urban soundscapes: {Experiences} and knowledge.
\newblock \emph{Cities 22}, 5 (2005), 339--350.

\bibitem[RLB03]{raimbault_ambient_2003}
\textsc{Raimbault M., Lavandier C., Bérengier M.}:
\newblock Ambient sound assessment of urban environments: field studies in two
  {French} cities.
\newblock \emph{Applied Acoustics 64}, 12 (2003), 1241--1256.

\bibitem[SCMD19]{szubert2019structure}
\textsc{Szubert B., Cole J.~E., Monaco C., Drozdov I.}:
\newblock Structure-preserving visualisation of high dimensional single-cell
  datasets.
\newblock \emph{Scientific reports 9}, 1 (2019), 1--10.

\bibitem[SL15]{seress2015habitat}
\textsc{Seress G., Liker A.}:
\newblock Habitat urbanization and its effects on birds.
\newblock \emph{Acta Zoologica Academiae Scientiarum Hungaricae 61}, 4 (2015),
  373--408.

\bibitem[Sla13]{slabbekoorn2013songs}
\textsc{Slabbekoorn H.}:
\newblock Songs of the city: noise-dependent spectral plasticity in the
  acoustic phenotype of urban birds.
\newblock \emph{Animal Behaviour 85}, 5 (2013), 1089--1099.

\bibitem[STN{\etalchar{*}}16]{smilkov_embedding_2016}
\textsc{Smilkov D., Thorat N., Nicholson C., Reif E., Viégas F.~B., Wattenberg
  M.}:
\newblock Embedding projector: {Interactive} visualization and interpretation
  of embeddings.
\newblock \emph{arXiv preprint arXiv:1611.05469} (2016).

\bibitem[tab07]{taber_technology_2007}
\emph{Technology for a quieter {America}, {National Academy of Engineering}}.
\newblock Tech. rep., Technical report, NAEPR-06-01-A, 2007.

\bibitem[TGdCQR20]{tagliasacchi_pre-training_2020}
\textsc{Tagliasacchi M., Gfeller B., de~Chaumont~Quitry F., Roblek D.}:
\newblock Pre-training audio representations with self-supervision.
\newblock \emph{IEEE Signal Processing Letters 27} (2020), 600--604.

\bibitem[VKDSVK14]{van_kempen_characterizing_2014}
\textsc{Van~Kempen E., Devilee J., Swart W., Van~Kamp I.}:
\newblock Characterizing urban areas with good sound quality: {Development} of
  a research protocol.
\newblock \emph{Noise and Health 16}, 73 (2014), 380.

\bibitem[{Was}21]{wspbirds}
\textsc{{Washington Square Park Eco Projects}}:
\newblock Explore birds, 2021.
\newblock URL:
  \url{https://www.wspecoprojects.org/our-projects/explore-birds/}.

\bibitem[WBS{\etalchar{*}}21]{wang2021calls}
\textsc{Wang Y., Bryan N.~J., Salamon J., Cartwright M., Bello J.~P.}:
\newblock Who calls the shots? {R}ethinking few-shot learning for audio.
\newblock In \emph{2021 IEEE Workshop on Applications of Signal Processing to
  Audio and Acoustics (WASPAA)} (2021), IEEE, pp.~36--40.

\bibitem[Wil21]{wilkinghoff_open-set_2021}
\textsc{Wilkinghoff K.}:
\newblock On open-set classification with {L3}-{Net} embeddings for machine
  listening applications.
\newblock In \emph{2020 28th {European} {Signal} {Processing} {Conference}
  ({EUSIPCO})} (2021), IEEE, pp.~800--804.

\bibitem[WLY{\etalchar{*}}13]{wang_visual_2013}
\textsc{Wang Z., Lu M., Yuan X., Zhang J., Van De~Wetering H.}:
\newblock Visual traffic jam analysis based on trajectory data.
\newblock \emph{IEEE Transactions on Visualization and Computer Graphics 19},
  12 (2013), 2159--2168.

\bibitem[WMCB19]{wang2019active}
\textsc{Wang Y., Mendez A. E.~M., Cartwright M., Bello J.~P.}:
\newblock Active learning for efficient audio annotation and classification
  with a large amount of unlabeled data.
\newblock In \emph{2019 {IEEE} International Conference on Acoustics, Speech
  and Signal Processing ({ICASSP})} (2019), IEEE, pp.~880--884.

\bibitem[Wys17]{wyse_audio_2017}
\textsc{Wyse L.}:
\newblock Audio spectrogram representations for processing with convolutional
  neural networks.
\newblock \emph{arXiv preprint ID:1706.09559} (2017).

\bibitem[XZ19]{XIE201974}
\textsc{Xie J., Zhu M.}:
\newblock Handcrafted features and late fusion with deep learning for bird
  sound classification.
\newblock \emph{Ecological Informatics 52} (2019), 74--81.

\bibitem[YCBL14]{yosinski_how_2014}
\textsc{Yosinski J., Clune J., Bengio Y., Lipson H.}:
\newblock How {Transferable} {Are} {Features} in {Deep} {Neural} {Networks}?
\newblock In \emph{Procedings of the 27th {International} {Conference} on
  {Neural} {Information} {Processing} {Systems} - {Volume} 2} (2014),
  {NIPS}'14, MIT Press, pp.~3320--3328.

\bibitem[YWL{\etalchar{*}}15]{yu2015iviztrans}
\textsc{Yu L., Wu W., Li X., Li G., Ng W.~S., Ng S.-K., Huang Z., Arunan A.,
  Watt H.~M.}:
\newblock iviztrans: Interactive visual learning for home and work place
  detection from massive public transportation data.
\newblock In \emph{2015 IEEE Conference on Visual Analytics Science and
  Technology (VAST)} (2015), IEEE, pp.~49--56.

\bibitem[ZFA{\etalchar{*}}14]{zeng_visualizing_2014}
\textsc{Zeng W., Fu C.-W., Arisona S.~M., Erath A., Qu H.}:
\newblock Visualizing mobility of public transportation system.
\newblock \emph{IEEE Transactions on Visualization and Computer Graphics 20},
  12 (2014), 1833--1842.

\bibitem[ZLH13]{zheng_u-air_2013}
\textsc{Zheng Y., Liu F., Hsieh H.-P.}:
\newblock U-air: {When} urban air quality inference meets big data.
\newblock In \emph{Procedings of the 19th {ACM} {SIGKDD} international
  conference on {Knowledge} discovery and data mining} (2013), pp.~1436--1444.

\bibitem[ZLW{\etalchar{*}}14]{zheng2014diagnosing}
\textsc{Zheng Y., Liu T., Wang Y., Zhu Y., Liu Y., Chang E.}:
\newblock Diagnosing new york city's noises with ubiquitous data.
\newblock In \emph{Proceedings of the 2014 ACM International Joint Conference
  on Pervasive and Ubiquitous Computing} (2014), pp.~715--725.

\bibitem[ZWC{\etalchar{*}}16]{zheng_visual_2016}
\textsc{Zheng Y., Wu W., Chen Y., Qu H., Ni L.~M.}:
\newblock Visual analytics in urban computing: {An} overview.
\newblock \emph{IEEE Transactions on Big Data 2}, 3 (2016), 276--296.

\bibitem[ZWVW20]{zahalka2020ii}
\textsc{Zah{\'a}lka J., Worring M., Van~Wijk J.~J.}:
\newblock Ii-20: Intelligent and pragmatic analytic categorization of image
  collections.
\newblock \emph{IEEE Transactions on Visualization and Computer Graphics}
  (2020).

\bibitem[ZYM{\etalchar{*}}14]{zhang_visual_2014}
\textsc{Zhang J., Yanli E., Ma J., Zhao Y., Xu B., Sun L., Chen J., Yuan X.}:
\newblock Visual analysis of public utility service problems in a metropolis.
\newblock \emph{IEEE Transactions on Visualization and Computer Graphics 20},
  12 (2014), 1843--1852.

\end{thebibliography}
